\documentclass[graybox]{svmult}

\usepackage{mathptmx}       
\usepackage{helvet}         
\usepackage{courier}        
\usepackage{graphicx}        

\usepackage[numbers]{natbib}
\usepackage{amssymb}

\def\arcmin{\hbox{$^\prime$}}
\def\arcsec{\hbox{$^{\prime\prime}$}}
\def\farcs{\hbox{$.\!\!^{\prime\prime}$}}

\begin{document}

\title*{3D Spectroscopic Observations of Star-Forming Dwarf Galaxies}
\author{Peter M.\ Weilbacher, Luz Marina Cair\'os, Nicola Caon, Polychronis Papaderos}
\institute{Peter M.\ Weilbacher \at Astrophysikalisches Institut Potsdam (AIP),
           An der Sternwarte 16, D-14482 Potsdam, Germany
           \email{pweilbacher@aip.de}
  \and Luz Marina Cair\'os \at Astrophysikalisches Institut Potsdam (AIP),
       An der Sternwarte 16, D-14482 Potsdam, Germany
  \and Nicola Caon \at Instituto de Astrof\'{i}sica de Canarias (IAC), C/ V\'ia L\'actea,
       s/n E38205 - La Laguna (Tenerife), Spain
  \and Polychronis Papaderos \at Centro de Astrof\'{i}sica da Universidade do Porto (CAUP),
       Rua das Estrelas 4150-762 Porto, Portugal
}
\maketitle

\abstract{We give an introduction into the observational technique of {\it
integral field} or {\it 3D} spectroscopy. We discuss advantages and drawbacks
of this type of observations and highlight a few science projects enabled by
this method.
In the second part we describe our 3D spectroscopic survey of Blue Compact
Dwarf Galaxies. We show preliminary results from data taken with the VIMOS
integral field unit and give an outlook on how automated spectral analysis and
forthcoming instruments can provide a new view on star formation
and associated processes in dwarf galaxies.
}


\section{Integral Field Spectroscopy}\label{sec:ifs}
Integral Field Spectroscopy (IFS) is a three-dimensional observing technique,
producing a spectrum for every sampled element on the sky. The end result after
data reduction is therefore typically a "datacube", with two spatial axes and
the spectral direction. This method is therefore often called {\it 3D Spectroscopy}.

\subsection{IFS basics}\label{sec:ifsbasics}
Unlike scanning techniques, an integral field unit (IFU) splits up the
telescope focal plane into a series of elements (typically lenslets, fibers, or
slicers, or a combination of them) and forms a pseudoslit from which the light
is dispersed to be recorded on a CCD. Such an exposure includes all wavelengths
and full spatial sampling and extent of the instrument, recorded at the same
time and under the same (atmospheric) conditions. Hence, IFS is a very
efficient observing scheme, especially if the objects of interest fit into the
field of view of one pointing. This is also one of the drawbacks of IFS: the
current instruments only cover a relatively small area on the sky, up to
$\sim$110\arcsec\ (as for VIRUS-P; \cite{HMS+08}). The other disadvantage is
that the information content of each exposure is extremely high, and depending
on the instrument, it can take a long time to properly (re-)construct the
datacube from the raw data.
One can therefore say that in general, IFS shifts the effort from the time
spent to carry out observing programs to the time required to do proper data
reduction (and data analysis). But recently, more and more software packages
are becoming available for IFS data reduction. Many instrument-specific
pipelines are being maintained (at e.g.\ the ESO and Gemini observatories) but
there are also more general reduction packages for fiber-based IFU data. The
newest package which has been adapted to reduce data from almost ten of the
most popular integral field instruments is {\sc p3d} \footnote{Available from
the project webpage {\tt http://p3d.sourceforge.net}.} \citep{SBR+10}.

Of course, IFS would not be rising in popularity, if there were not advantages
that by far outweigh the difficulties, at least for selected scientific
applications. Obviously, it enables bidimensional spectral coverage (as opposed
to traditional slit spectroscopy) with all the advantages that implies: no slit
effects or slit losses, no necessity to do (pre-)imaging of the same field, a
spectral resolution that is independent of spatial coverage, possibility to
correct for atmospheric dispersion and to bin the data spatially to gain S/N.
IFS also has potentially long wavelength coverage (as compared to the
Fabry-Perot technique) and, as said above, guarantees homogeneous conditions
for the whole dataset (as opposed to scanning techniques).

\subsection{IFS advantage: no slit effects}\label{sec:sliteffects}
The problem of slit effects \citep[see e.g.][]{BAB+95} for a detailed
discussion) is particularly troublesome for slit spectroscopy, which many
astronomers are not aware of. If one observes one or more compact objects with
a slit spectrograph under seeing conditions which cause a FWHM smaller than the
slit width, velocities measured for each object depend on their centering
within the slit. This can be corrected for stellar fields with precisely known
star positions (as done by \cite{GvdM+02}). But for emission line kinematics of
compact objects such a correction is not possible, unless a narrow-band image
with sufficient spatial resolution exists, from which the exact positions of
all relevant emission line peaks can be determined. Emission line kinematics of
e.g.\ star-forming dwarf galaxies derived using long-slits is therefore to be
regarded with caution.

To give an example: if one observes H$\alpha$ with a spectral resolution of
$R=500$ at slit width of 1\arcsec, the slit is effectively $\sim
600$\,km\,s$^{-1}$ wide. At a resolution of $R=10000$ it still is $\sim
30$\,km\,s$^{-1}$ wide. So, if one happens to observe a compact HII region in
0\farcs5 seeing with a 1\arcsec\ slit and $R=10000$, one on average makes an
error of $\pm$15\,km\,s$^{-1}$ (for shifts of half the slit width the FWHM of
the object is still within the slit) at each position along the slit. Rotation
curves derived that way can look very much like disturbances caused by e.g.\
galaxy merging and lead to wrong interpretation of the galaxy at hand.

Most kinds of integral field spectrographs are insensitive to slit effects:
instruments which use lenslets are insusceptible to this problem; fiber-based
systems are less affected as fibers scramble the spatial signal sufficiently
well. Only IFUs built using slicers with low spatial sampling can be affected.
But then one can derive the spatial position of most objects within the field
of view, allowing to correct for the instrumental shifts. Of course,
instruments based on the Fabry-Perot principle are also immune. To derive
emisson line kinematics of any object one should therefore always choose to
observe with an integral field spectrograph or a similar instrument that is not
affected by slit effects.

\subsection{Analyses enabled by IFS}
Some kinds of analyses are only possible if one observes galaxies with IFS.
Basically, one can create two-dimensional maps of many properties, which is not
possible using slit spectroscopy. One can even create spatially resolved maps
involving line ratios or equivalent widths (EWs) which is possible with
Fabry-Perot instruments but expensive in terms of observing time.
Typical IFS-based maps are those for extinction (Balmer decrement) and
chemical abundances.
Two-dimensional spectroscopy also allows to create color maps from which the
effects of dust and contributions by emission line fluxes were removed; these
can then be interpreted using photometric stellar population models without the
need to build complex models involving extinction and line emission.
Kinematical analyses are of course possible, but can be done using other
observing methods. Another unique analysis enabled by IFU-observations is the
search for spectral signatures across a galaxy. This is typically done for
Wolf-Rayet features in star-forming objects.

Projects like SAURON (kinematics and stellar populations of nearby elliptical
galaxies; \cite{BCM+01,ECK+07}) and SINS (properties of $z\sim2$ galaxies;
\cite{FSG+06,FSG+09}) would not have been possible without integral field
spectroscopy (and a well-managed team working on data reduction and analysis).

\section{Studies of Blue Compact Dwarf galaxies with IFS}\label{sec:BCDs}
Among the many recent publications involving integral field spectroscopy of
dwarf galaxies we want to highlight those of an ongoing, comprehensive study of
Blue Compact Dwarf (BCD) galaxies. BCDs are dwarf starburst galaxies with low
metallicity, often regarded as local counterparts of young galaxies observed at
high-redshift. Although it became clear later on that the vast majority of
these dwarf galaxies have an underlying old stellar population
\cite{P96,CVGP+01}, these objects are still the best nearby laboratories for
exploring violent star formation and dwarf galaxy build up, including the
feedback processes associated with it.
Some among the many open questions which we try to address using IFS:
What are the star-formation histories of BCDs?
What is their dust content and distribution?
Are there evolutionary connections between star-forming dwarfs and other types of dwarfs?
What mechanisms trigger the starburst?
Are interactions and mergers playing a role?

\subsection{Project description}
In our project (PI L.~M.~Cair\'os) we aim to thoroughly analyze a large sample of
$\gtrsim$40 BCDs using integral field spectroscopy.
Our targets are located in both hemispheres and cover a wide range in
luminosity ($-13.9 \lesssim M_B \lesssim -21.1$\, mag).
We observed BCDs over a wide metallicity ($0.05 < Z_\odot < 0.85$) and
morphology range \citep{CVGP+01}. Deep (optical) imaging in at least two bands
is available for all our targets. By now we have acquired data from four
instruments/telescopes: INTEGRAL (coupled to the WYFFOS spectrograph at WHT),
PMAS (3.5m Calar Alto telescope), VIRUS-P (at the 2.7m McDonald telescope), and
VIMOS (at the ESO VLT).

Our methodology involves a thorough analysis of the data of each object. We
first build maps of emission line fluxes or equivalent widths for different
line types (hydrogen Balmer lines, forbidden oxygen or sulfur lines, etc.).
These maps allow us to localize the HII regions and investigate the luminosity,
morphology, excitation mechanisms, extinction pattern, and kinematics of the
ionized gas component.  From maps of the continuum we can study the morphology
of the underlying stellar component, undisturbed by line emission, and better
assess its age structure. The Balmer decrement enables us to carry out a
spatially resolved study of the intrinsic extinction.  Further insights can be
obtained from Wolf-Rayet features (based on the study of the emission lines in
the spectral range 4650\dots4690\,\AA) which gives clues to the starburst age
and the shape of the initial mass function.  Metal emission lines can tell us
about physical parameters of the gas and spatial variations of the chemical
abundances.  Using gas and stellar velocity fields we investigate the origin of
starburst activity and estimate the total mass of the galaxy.  Finally, we use
spatially resolved spectral synthesis models to characterize the properties of
stellar populations in the sample galaxies (see \ref{sec:auto}).

Some of our data taken with INTEGRAL and PMAS has already been published. An
initial investigation of five galaxies with INTEGRAL was presented in
\cite{GCC+08}, and a paper first showing the potential of the PMAS data for
the case of Mrk\,1418 appeared as \cite{CCZ+09}. The PMAS data on Mrk\,409 was
thoroughly exploited by \cite{CCP+09} and a general presentation of the
objects observed so far with PMAS was published as \cite{CCZ+10}.
A further description of the sub-sample of BCDs observed with the PMAS and VIRUS-P
instruments can also be found in \citep[][this volume]{Cairos11}.
All these papers show results that would have been impossible to acquire with
other observing techniques.

\subsection{VIMOS data}
Nine southern galaxies of our sample were so far observed with the VIMOS
spectrograph on the 8.2m ESO VLT. This dataset contains our deepest exposures
with the highest spectral resolution ($R\sim2600$). We used the HRblue and
HRorange grisms to cover most of the optical wavelength range from 4300 to
7400\,\AA. We used the setup with low spatial resolution
(0.67\arcsec\,px$^{-1}$) that was well matched to the seeing and allowed to
cover $\sim27\arcsec\times27\arcsec$ per target. Our observing strategy involved
several dithered exposures per galaxy and grism, to fill in bad spatial elements
(spaxels) and to aid sky subtraction. This increased the complexity of the data
reduction, since the combination of multiple dithered exposures into a full
datacube is not supported by any existing processing tool.

\begin{figure}
\includegraphics[width=\linewidth,height=0.5\linewidth]{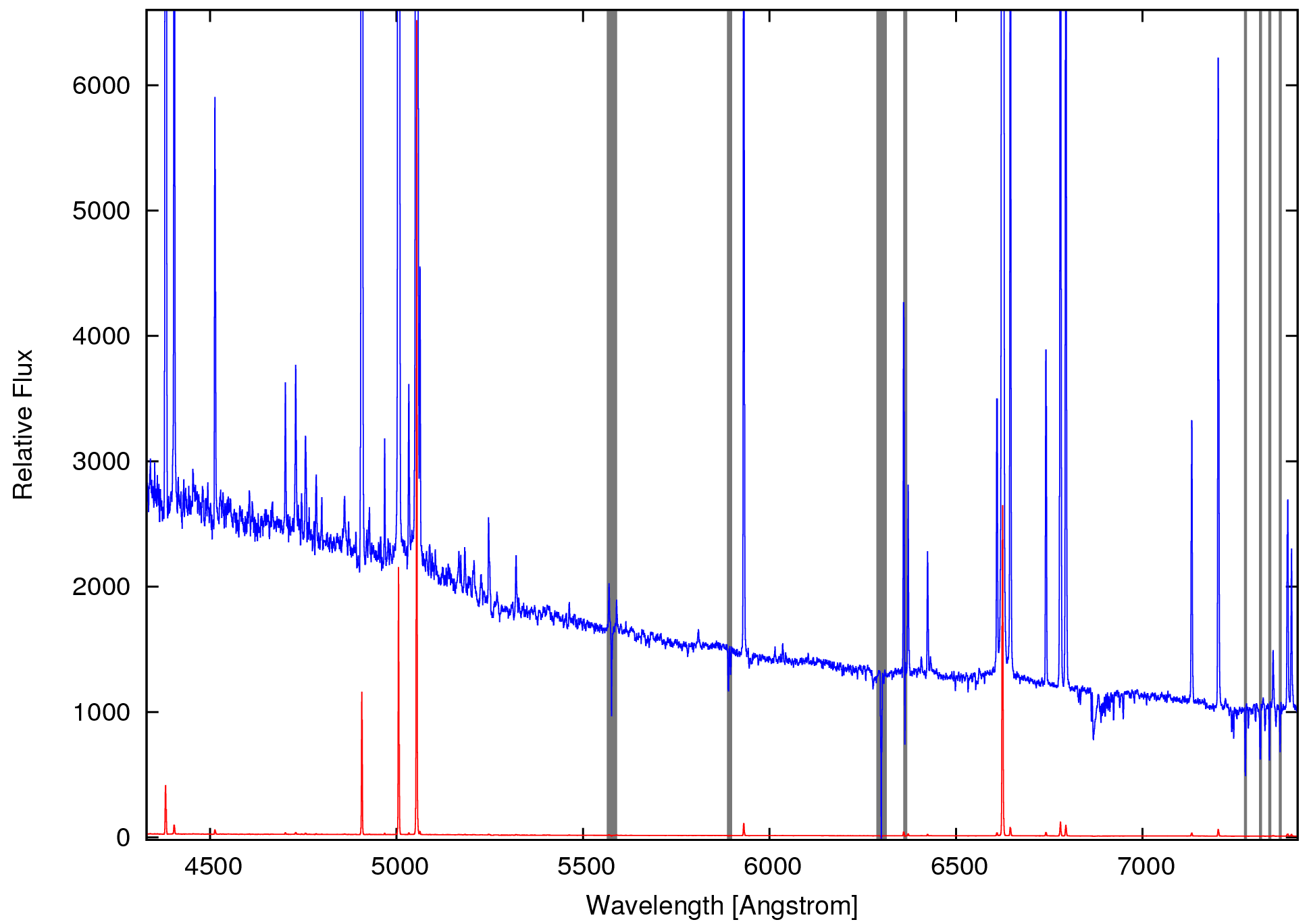}
\caption{Spectrum of Tol\, 1924-416: this spectrum was integrated over the 80
         central spectral elements of our VIMOS data. It shows many faint He and
         metal lines with detections at good S/N. The lower spectrum is the same
         as the upper one but scaled down 100$\times$ to show the bright emission
         lines. Grey vertical semitransparent areas indicate sky emission lines
         that could not be subtracted perfectly.}\label{fig:Tol1924_Spec}
\end{figure}

\begin{figure}
\includegraphics[width=0.5\linewidth]{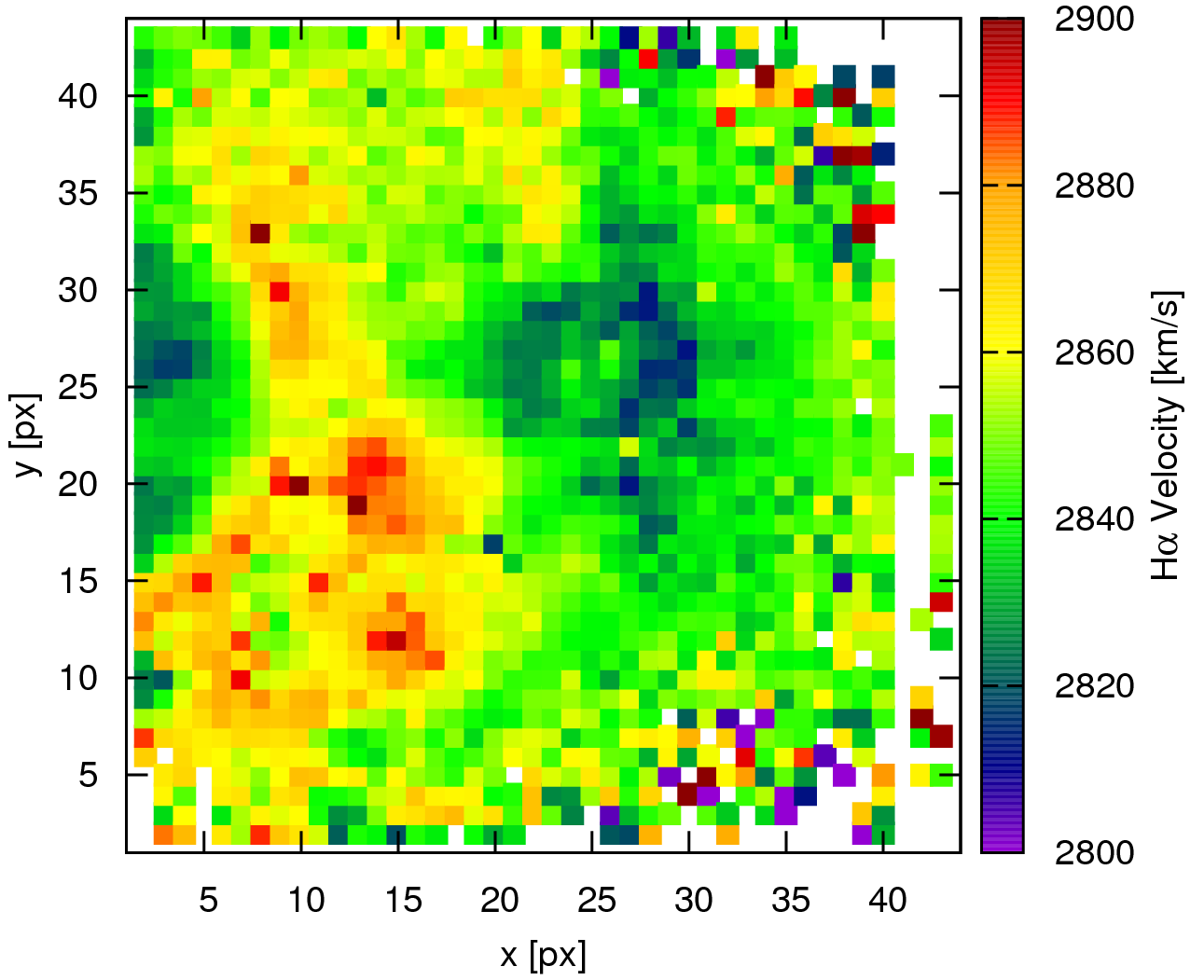}
\includegraphics[width=0.5\linewidth]{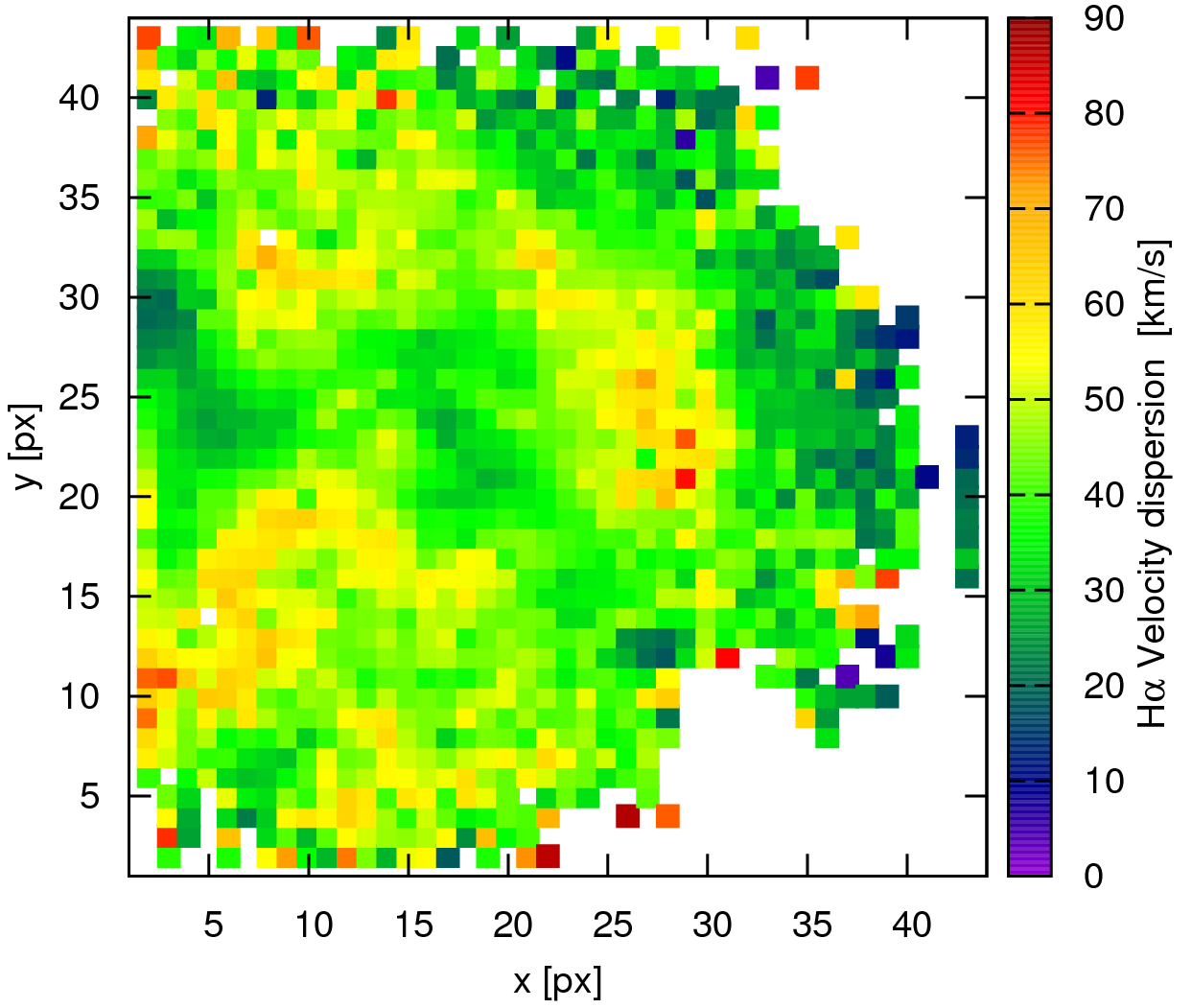}
\caption{Kinematical maps of Tol\, 1924-416: {\it left} is the velocity field
         as derived from H$\alpha$, {\it right} is the velocity dispersion
         measured from the same line.}\label{fig:Tol1924_Map}
\end{figure}

Here, we show preliminary results of one object of our VIMOS sample,
Tol\,1924-416. Figure~\ref{fig:Tol1924_Spec} shows an integrated spectrum of
the central part of Tol\,1924-416. This high S/N spectrum shows many faint
Helium and metal lines that are not commonly seen in BCDs.
Figure~\ref{fig:Tol1924_Map} presents maps of velocity and velocity dispersion
as derived from the H$\alpha$ emission line.  While the emission line
morphology is similar to the broad-band appearance, the velocity structure is
very complex.  While local starburst-driven winds might contribute to this
structure, it is tempting to take it as a sign of merging,
similar to what previous investigations of this galaxy concluded from optical
broad- and narrow-band observations \citep{HV95,DCC99}.  Derivation of stellar
velocities in this galaxy, to confirm or disprove this view, are difficult as
only very few spaxels show absorption.

\section{Conclusions \& Outlook}\label{sec:out}
In this article we described the technique of integral field spectroscopy that
provides an unbiased and unrestricted view on dwarf galaxies: from the same
data, taken under nearly identical conditions, one can derive kinematics
(gaseous and possibly stellar) and physical properties of the warm interstellar
medium (e.g.\ metallicity, density), and study the stellar populations in two
dimensions.
We briefly outlined a project investigating the structure and evolution of BCDs
with IFS and showed preliminary results for one galaxy from our sample.

But most of the findings enabled by 3D spectroscopy are probably still to come.
Next, we briefly discuss two topics that will help to further exploit existing data
and to produce major new results, in the research area of dwarf galaxies and
beyond.

\subsection{Automated spectral analysis}\label{sec:auto}
\begin{figure}
\includegraphics[width=\linewidth]{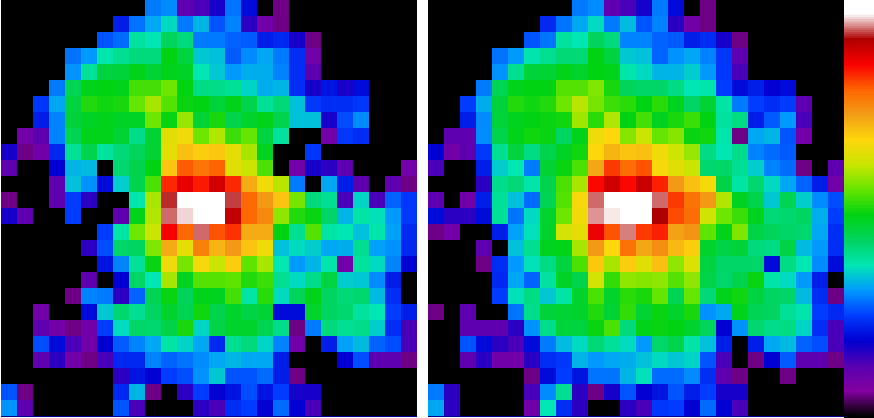}
\caption{Automated stellar continuum analysis of VIRUS-P data {for the BCD}
  III\,Zw\,102: the {\it left} and {\it right} panels show, respectively, the
  H$\beta$ map of the galaxy prior to and after correction for underlying
  stellar absorption.}
\label{fig:IIIZw102}
\end{figure}

Most IFS-based investigations so far analyze the integrated spectrum of
multiple regions of interest, or perform Gaussian emission line fitting to
individual spaxels, leaving unexploited the information provided by stellar
absorption lines.  As deeper data gets available, the latter will become more
accessible and can be used, especially for kinematical studies.

Another approach is to use spectral synthesis and analysis tools to model the
stellar continuum spaxel by spaxel. This yields a twofold benefit, as it allows
for insights into the stellar populations and for the correction for underlying
stellar absorption, thereby an improvement on the accuracy of emission line
measurements. Especially Balmer line fluxes usually have to be corrected for
underlying stellar absorption. The subtraction of the best-fitting stellar
model offers an effective means for accomplishing this task.

As an example, we show in Fig.~\ref{fig:IIIZw102} preliminary results from the
analysis of our VIRUS-P data for the BCD III\,Zw\,102 (Cair\'os et al., in
prep.). We used the STARLIGHT code \citep{CFG+04,CFM+05} to generate a stellar
continuum fit at each position. {Figure}~\ref{fig:IIIZw102} shows the H$\beta$
flux before and after correction for underlying stellar absorption using the
best-fitting stellar model for each spaxel. It can be seen that the artifical
depression in the H$\beta$ flux distribution in southeastern and northwestern
direction could be efficiently rectified this way (Papaderos et al., in prep.).

\subsection{New instruments}
Several major new IFSs are scheduled to come on-line in the next years. At the
ESO VLT, all new instruments will be based on the integral field principle or
will at least contain an ancillary IFU-mode. The major IFSs to become available at
ESO are KMOS, a near-infrared spectrograph with 24 small IFUs, and MUSE, an
optical instrument. The VIRUS instrument in development for the HET at McDonald
observatory will provide simultanous (but not contiguous) spectral coverage of
an area of 30\,$\square$\arcmin\ using a setup of up to $\sim150$ fiber bundles.

Of these, the {\bf M}ulti {\bf U}nit {\bf S}pectroscopic {\bf E}xplorer
(MUSE; \cite{Bacon+10}) will probably be the most powerful instrument to
further drive ahead the field of nearby star-forming dwarf galaxies. It is
currently being built as one of the 2nd generation instruments for the ESO VLT
and will see first light in 2012. Its 1\arcmin$\times$1\arcmin\ field of view at
0\farcs2 sampling, long wavelength range of (4650) 4800$\dots$9300\,\AA,
relatively high spectral resolution ($R\sim3000$), and high throughput should
make it an ideal multi-purpose instrument. After first light, it will be
enhanced with an adaptive optics module to in the end allow near-diffraction
limited spectroscopy in a 7\farcs5$\times$7\farcs5 field with a 0\farcs025
sampling.

\begin{acknowledgement}
PMW acknowledges support by the German Verbundforschung through the MUSE/D3Dnet
project (grant 05A08BA1).
PP is supported by a Ciencia 2008 contract, funded by FCT/MCTES (Portugal) and
POPH/FSE (EC).

\end{acknowledgement}

\end{document}